\documentclass[structabstract]{aa}  
\usepackage{graphicx}
\usepackage{txfonts}
\usepackage{natbib}
\bibpunct{(}{)}{;}{a}{}{,}
\begin{document}
   \title{Thermohaline mixing and the photospheric composition of low-mass giant stars}


   \author{F. C. Wachlin
          \and
           M. M. Miller Bertolami
          \and
           L. G. Althaus
          }

   \institute{Facultad de Ciencias Astron\'omicas y Geof\'{\i}sicas,
              Universidad Nacional de La Plata, Argentina\\
             Instituto de Astrof\'{\i}sica de La Plata, CONICET La Plata -- UNLP, Argentina\\
             \email{fcw@fcaglp.unlp.edu.ar}
             }


 
  \abstract
   {}
   {We test by means of numerical simulations and different recipes the efficiency of thermohaline mixing as a process to alter the surface abundances in low-mass giant stars.}
   {We compute full evolutionary sequences of red giant branch stars close to the luminosity bump by including state of the art composition transport prescriptions for the thermohaline mixing regimes. In particular we adopt a self-consistent double-diffusive convection theory, that allows to handle the instabilities that arise when thermal and composition gradients compete against each other, and a very recent empirically motivated and parameter free asymptotic scaling law for thermohaline composition transport.}
   {In agreement with previous works, we find that during the red giant stage, a thermohaline instability sets in shortly after the hydrogen burning shell (HBS) encounters the chemical discontinuity left behind by the first dredge-up. We also find that the thermohaline unstable region, initially appearing at the exterior wing of the HBS, is unable to reach the outer convective envelope, with the consequence that no mixing of elements that produces a non-canonical modification of the stellar surface abundances occurs. Also in agreement with previous works, we find that by artificially increasing the mixing efficiency of thermohaline regions it is possible to connect both unstable regions, thus affecting the photospheric composition. However, we find that in order to reproduce the observed abundances of red giant branch stars close to the luminosity bump, thermohaline mixing efficiency has to be artificially increased by about 4 orders of magnitude from that predicted by recent 3D numerical simulations of thermohaline convection close to astrophysical environments. From this we conclude the chemical abundance anomalies of red giant stars cannot be explained on the basis of thermohaline mixing alone.}
   {}

 \titlerunning{Thermohaline mixing in low-mass giant stars}
   \maketitle
%
%
\section{Introduction}
After leaving the main sequence, low-mass stars move in the HR diagram towards the red giant branch (RGB). During the RGB, nuclear reactions take place in a thin shell surrounding the helium core and moving outwards in mass. The material processed by H-burning is kept hidden inside the core until the inner boundary of the convective envelope penetrates deeply inwards, reaching the freshly synthetised nucleides. When this happens, the material processed by nuclear reactions is dredged up to the surface (in the so called first dredge up) modifying the photospheric composition of red giant stars. Standard stellar evolution theory \citep{iben67} predicts no further surface abundance variation should take place. However, observational evidence strongly suggests the existence of a non-canonical mixing processes on the RGB \citep{gilroy89, gilroy-brown91, luck94, charbonnel94, charbonnel-brown-wallerstein98, charbonnel-donascimento98, gratton-sneden-carretta-bragaglia00, smith-hinkle-et-al02,  shetrone03, geisler-smith-et-al05, spite-cayrel-et-al06, recio-blanco-laverny07, smiljanic-gauderon-et-al09}. This extra-mixing seems to be related to the RGB luminosity-function bump, i.e., the phase of the evolution when the narrow hydrogen burning shell reaches the chemical discontinuity caused by the deep penetration of the convective envelope, leading to a transitory drop of the luminosity of the star and producing a peak in the giant branch luminosity distribution.

In the last years considerable effort has been devoted to identify the non-canonical physical processes that could be responsible for modifying the photospheric composition of low-mass giant stars at the luminosity bump stage. One important clue was first provided by \citet{eggleton-dearborn-lattanzio06} by detecting the appearence of a mean molecular weight ($\mu$) inversion in a region just above the HBS when the burning shell reached the uniform composition layers left behind by the first dredge-up phase. Using the classic Rayleigh-Taylor criterion, \citet{eggleton-dearborn-lattanzio06} found this region to be hydrodynamically unstable. The $\mu$-inversion detected was identified to come from the $^3$He($^3$He,2p)$^4$He reaction, a process that takes two nuclei and transforms them into three, producing a local depression in the mean molecular weight per nuclei. This depression is very tiny and becomes evident just when it takes place in a background of homogeneous chemical composition, like that found by the external wing of the HBS at the luminosity bump region.

\citet[CZ07]{charbonnel-zahn07} pointed out that, in a star, rather than a dynamical instability (Rayleigh-Taylor) it is a double-diffusive instability (known in the literature under the name of thermohaline instability) what first sets in as the inverse $\mu$-gradient builds up. This thermohaline instability takes place when the stabilizing agent (heat) diffuses away faster than the destabilizing agent ($\mu$), leading to a slow mixing process that might provide the extra-mixing seeked. 

Since thermohaline instability was identified to take place at the luminosity bump several efforts have been conducted in order to understand the actual relevance of this process in modifying the surface abundance composition of low mass giant stars along the RGB \citep{denissenkov-pinsonneault08, cantiello-langer10, charbonnel-lagarde10, denissenkov10, denissenkov-merryfield11, stancliffe10}. So, the thermohaline mixing has been studied by means of numerical simulations either by considering it as an isolated process or combining it with other mechanisms which might contribute to the turbulence of the material (e.g., rotation, internal magnetic fields). 

Previous investigations are uncertain about the consequences and relevance of the setting in of thermohaline instability at the RGB: while some authors find the mixing rate generated by this process to be enough to reproduce the surface abundances observed, others encounter this mechanism to be insufficient and propose the interaction of more than one process to explain the obsevations. Surely a realistic scenario should take into account all possible physical processes present and study how they contribute and interact among themselves. However, there are still important doubts in the current treatment of the thermohaline mixing as an isolated process which should be addressed before we consider actually understood the role of this mechanism.

One important source of uncertainty comes from the calibration of the degree of turbulence generated by each instability. In particular, the thermohaline instability gives rise to a slow mixing of the material which is usually treated as a diffusive process characterized by a coefficient that determines the efficiency of the mixing. This parameter, the diffusion coefficient, has to be set beforehand in order to solve the corresponding diffusion equation. To this end, it has usually been adopted a prescription based on the work of \citet{ulrich72} and \citet{kippenhahn-ruschenplatt-thomas80}, where the diffusion coefficient was found to be proportional to the square of the (unknown) aspect ratio $\alpha$ (length/width) of fluid elements. Unfortunately this means a great uncertainty in its value since the linear theory does not give a reliable estimate of the maximum length of salt fingers relative to their diameter. 

Being the subject still a matter of debate, laboratory experiments simulating oceanic conditions \citep[e.g.,][]{krishnamurti03} suggest a geometry of slender fingers for the convective elements and thus some authors adopt high values of $\alpha$ ($\geq 5$) in order to reproduce the surface abundances of low-mass stars after the luminosity bump. However, physical conditions inside a star are very different from those in the laboratory and it is not clear if elongated structures can be stable, especially when shear and horizontal turbulence is present. In view of these concerns, other authors \citep[e.g.,][]{kippenhahn-ruschenplatt-thomas80, cantiello-langer10} adopt "blobs" ($\alpha\approx 1$) as the preferred fluid element morphology. This freedom in the election of the aspect ratio has an evident impact on the diffusion coefficient which has been reported to affect the results \citep[CZ07,][]{cantiello-langer10}.

Very recently, \citet{denissenkov10} and \citet{traxler-garaud-stellmach11} have presented the first numerical simulations of thermohaline (fingering) convection close to the astrophysical regime --- i.e. Prandtl number $Pr\sim 10^{-6}$ and inverse Lewis number $\tau\sim 10^{-6}$. In fact, \citet{traxler-garaud-stellmach11}, by means of high performance three dimensional simulations, were able to derive asymptotic scaling laws for thermohaline composition transport. These asymptotic scaling laws are, then, the first empirically motivated and parameter free available recipe for the treatment of thermohaline mixing in an astrophysical regime. Both \citet{denissenkov10} and \citet{traxler-garaud-stellmach11} have suggested that their results imply that the thermohaline mixing is not efficient enough to account for the changes in the surface abundances of red giants close to the luminosity bump, but no stellar evolution computations have been performed. 

In the present work we test these suggestions by means of full evolutionary simulations of the development of thermohaline convection in RGB stars.
Specifically we study the relevance of thermohaline mixing in RGB stars when more sophisticated and physically sounding prescriptions, than that of \citet{kippenhahn-ruschenplatt-thomas80}, are adopted in a stellar evolutionary code. In particular we adopt the very recent prescription of \citet{traxler-garaud-stellmach11} and the double diffusive mixing length theory of \citet{grossman-taam96}. While the former is based on realistic 3D numerical experiments and is essentially parameter free, the latter successfully reproduces most previously known results about convection in astrophysics and, when composition gradients are considered, it establishes its own stability conditions, thus providing a new perspective to study the thermohaline instability problem. 

%
\section{GNA convection theory}
\label{sec:gnatheory}
As an effort to provide a better nonlocal theory of convection, \citet{grossman-narayan-arnett93} developed a flexible and powerful formalism, which was designed to make unbiased, self-consistent predictions about complex phenomena associated to the transport of energy in stars. Here we use this formalism and follow the prescription of \citet{grossman-taam96} to get the local theory of convection in a composition-stratified fluid.  

\begin{figure}
\resizebox{\hsize}{!}{\includegraphics{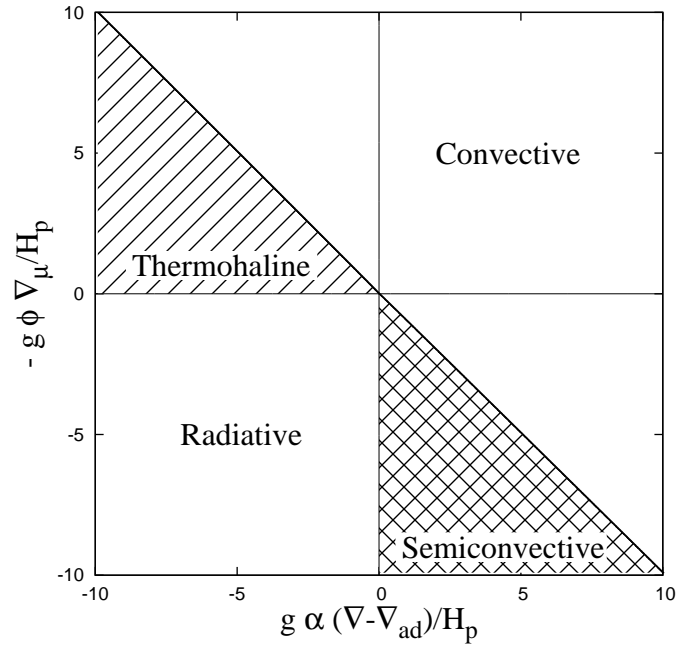}}
\caption{Regions of different stability regimes for diffusion rates $A=0$, $D=0.001$ and $F=0$. Note the convective region with $\nabla-\nabla_\mathrm{ad}<0$ for which the standard mixing length approach ($\nabla_{\mu}=0$) can not provide mixing velocities and that have been sometimes misidentified with the thermohaline regime.}
\label{fig:stab-regions}
\end{figure}

Basically the theory allows to find the mixing rate of the fluid in the convective, thermohaline and semiconvective regimes by solving two equations simultaneously: the first of these equation corresponds to the turbulent velocity $\sigma$,
\begin{eqnarray}
& \sigma^2 & \left[(A+D+2B\sigma){{g \alpha}\over{H_\mathrm{p}}} (\nabla-\nabla_\mathrm{ad})
- (A+F+2B\sigma) {{g \phi}\over{H_\mathrm{p}}} \nabla_\mathrm{\mu} \right. \nonumber \\
& & - (A+D+2B\sigma)(A+F+2B\sigma)(D+F+2B\sigma)\biggr] \nonumber \\
& \times & \left[(F+B\sigma){{g \alpha}\over{H_\mathrm{p}}} (\nabla-\nabla_\mathrm{ad})
-(D+B\sigma){{g \phi}\over{H_\mathrm{p}}} \nabla_\mathrm{\mu} \right. \nonumber \\
& & -(A+B\sigma)(D+B\sigma)(F+B\sigma)\biggr] =0,
\label{eqnsigma}
\end{eqnarray}
where $\nabla=\partial \ln T/\partial \ln P$, $\nabla_\mathrm{ad}=(\partial \ln T/\partial \ln P)_\mathrm{ad}$ is the adiabatic gradient, $\nabla_\mathrm{\mu}=\partial \ln \mu/\partial \ln P$ is the molecular weight gradient, $g$ is the local acceleration due to gravity, 
$\alpha=-(\partial \ln\rho/\partial\ln T)_\mathrm{P,\mu}$ is the coefficient of thermal expansion, 
$ \phi=(\partial \ln\rho/\partial\ln \mu)_\mathrm{P,T}$, $H_\mathrm{p}$ is the pressure scaleheight,  $D$, $F$ and $A$ are the diffusion rates of heat, composition and momentum, respectively,
and $B=2/{l}$, with $l$ the unique mixing length considered by \citet{grossman-taam96}. The other equation involves the flux conservation 
\begin{equation}
\nabla_\mathrm{Rad}-\nabla_\mathrm{ad} = (\nabla-\nabla_\mathrm{ad})+
H_\mathrm{p} \left({{\rho C_P}\over{ K T}}\right)~ \overline{w\theta}, 
\label{eqnflux}
\end{equation}
\begin{figure}
\resizebox{\hsize}{!}{\includegraphics{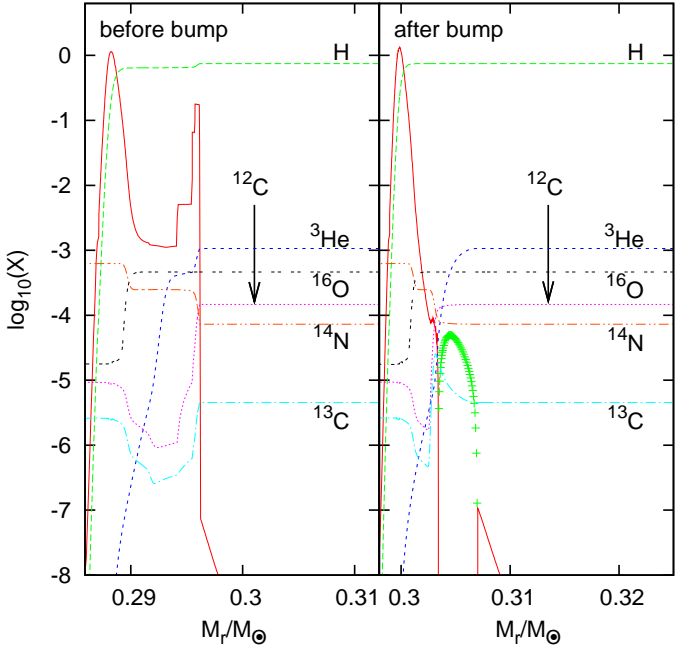}}
\caption{Profiles of the abundances of H, $^3$He, $^{12}$C, $^{13}$C, $^{14}$N, $^{16}$O and of the mean molecular weight gradient $\nabla_\mathrm{\mu}$ as a function of mass coordinate. The full line stands for $\nabla_\mathrm{\mu}>0$ and the plus signs otherwise. Left and right panels correspond to the situation before and after the bump, respectively, for a $0.9 M_{\sun}$ model with $1\times 10^{-3}$. The abscissa ranges from the bottom of the hydrogen burning shell to the base of the convective envelope.} 
\label{figure2}
\end{figure}
where $\nabla_\mathrm{Rad}$ is the temperature gradient that would be necessary to transport the whole flux by radiation, $\rho$ is the density, $C_P$ is the specific heat, $K$ is the radiative diffusive conductivity, and $\overline{w\theta}$ is the correlation between turbulent velocity and turbulent temperature excess given by
\begin{equation}
\overline{w\theta}=
{{
   \left( {X\over{D+F+2B\sigma}}-A-F-2B\sigma-{Y\over{F+B\sigma}}\right) 
   (A+B\sigma) - Y
}
\over{
   {X\over{D+F+2B\sigma}}-A-F-2B\sigma-{Y\over{F+B\sigma}}+Y
}}{{T\sigma^2}\over{\alpha g}}
\label{eq:wtheta}
\end{equation}
where, for simplicity, we have set that $X=g\alpha (\nabla-\nabla_\mathrm{ad})/H_\mathrm{p}$ and $Y=g\phi \nabla_\mathrm{\mu}/H_\mathrm{p}$. Our Eq. (\ref{eq:wtheta}) differs from Grossman \& Taam's Eq. (17) because we have fixed some sign errors present in the original expression. It is worth mentioning that for realistic cases $D\gg A$, $D\gg F$ and $D$ goes to zero.
 
Equations (\ref{eqnsigma}) and (\ref{eqnflux}) have to be solved simultaneously for $\sigma$ and $(\nabla-\nabla_\mathrm{ad})$. The leading factor of Eq. (\ref{eqnsigma}) shows that $\sigma=0$ is always a solution, but it corresponds to a stable equilibrium only if no other real and non-negative root exists. In general, the fluid will seek out the most turbulent equilibrium state, thus, if more than one root is positive the system will evolve to the largest root, being the $\sigma=0$ solution unstable. 

In the present work the mixing of nuclear species of mass fraction $X_i$ is performed by solving the diffusion equation 
\begin{equation}
{{d X_i} \over {dt}}=
\left( {{\partial X_i}\over{\partial t}} \right)_\mathrm{nuc}+
{\partial\over{\partial M_r}}
\left[ (4\pi r^2\rho)^2 D_\mathrm{c}{{\partial X_i}\over{\partial M_r}}\right]
\end{equation}
with the diffusion coefficient $D_\mathrm{c}$ defined in terms of the turbulent velocity $\sigma$ and the mixing length $l$ by \citep{weaver-zimmerman-woosley78}
\begin{equation}
D_\mathrm{c}={1\over 3}\sigma l.
\label{eq:diffcoef}
\end{equation}
Appendix A contains a few additional details about the procedure followed by us in order to solve GNA's equations.

%
\section{An empirical scaling law for compositional transport by fingering convection}
\label{sec:traxler}
Although double-diffusive processes have been studied by several 
 authors by means of hydrodynamics codes \citep[see, e.g.,][]{merryfield95,biello01,bascoul07,zaussinger-spruit10}, it was
\citet{traxler-garaud-stellmach11} who performed the first three-dimensional simulations to address the question of double-diffusive transport by fingering convection in astrophysics. It is important to note that \citet{traxler-garaud-stellmach11} conducted their simulations at Pr $\sim$ O($10^{-2}$) while the true astrophysical regime occurs at Pr $\sim$ O($10^{-6}$). Therefore, their empirical scaling law relies on the validity of the asymptotic behavior suggested by their results.
They model a finger-unstable region using a local Cartesian frame $(x,y,z)$ oriented so that its vertical axis $z$ has a direction opposite to that of the gravitational acceleration. Also the Boussinesq approximation is used and, consequently, it is assumed that small density, temperature and compositional perturbations $(\tilde\rho, \tilde T, \tilde\mu)$ are related by the following linearized equation
\begin{equation}
\frac{\tilde\rho}{\rho_0}=\alpha \tilde T + \beta \tilde \mu,
\label{eq:boussinesq}
\end{equation}
where $\rho_0$ is a reference density, $\alpha=-\rho_0^{-1} \partial\rho/\partial T$, and $\beta=\rho_0^{-1} \partial\rho/\partial\mu$.
Expressing the velocity, temperature and compositional fields as a background component plus a perturbation, it is obtained
\begin{equation}
{\bf u}(x,y,z,t)= \tilde {\bf u}(x,y,z,t),
\end{equation}
\begin{equation}
T(x,y,z,t)=T_0(z)+\tilde T(x,y,z,t),
\end{equation}
\begin{equation}
\mu(x,y,z,t)=\mu_0(z)+\tilde\mu(x,y,z,t),
\end{equation}
with $T_0(z)=z\, \partial T/\partial z$ and $\mu_0=z\, \partial \mu/\partial z$. By scaling the time ($t$), the temperature and the composition adequately by means of the expected finger scale \citep[see][for details]{traxler-garaud-stellmach11}, the final set of equations to solve turns out to be
\begin{equation}
\frac{1}{\mathrm{Pr}}\left( \frac{\partial\tilde {\bf u}}{\partial t} + \tilde{\bf u} \cdot \nabla\tilde {\bf u}\right) = -\nabla\tilde p + (\tilde T- \tilde\mu) \mathrm{\bf e}_z+\nabla^2 \tilde {\bf u},
\label{eq:trax6}
\end{equation}
\begin{equation}
\nabla \cdot\tilde {\bf u} = 0, 
\label{eq:trax7}
\end{equation}
\begin{equation}
\frac{\partial\tilde T}{\partial t}+ \tilde w + {\bf\tilde u}\cdot\nabla\tilde T =\nabla^2 \tilde T,
\label{eq:trax8}
\end{equation}
\begin{equation}
\frac{\partial\tilde\mu}{\partial t}+\frac{\tilde w}{R_0} + \tilde{\bf u}\cdot\nabla\tilde \mu=\tau\nabla^2\tilde\mu,
\label{eq:trax9}
\end{equation}
where $\tilde w$ is the $z$ component of $\tilde {\bf u}$, Pr is the Prandtl number, $\tilde p$ is the non-dimensional pressure perturbation from hydrostatic equilibrium, $R_0= (\nabla - \nabla_\mathrm{ad})/\nabla_\mu$ and $\tau=\kappa_\mu/\kappa_T$, with $\kappa_\mu$ the compositional diffusivity (see below) and $\kappa_T$ the thermal diffusivity given by
\begin{equation}
\kappa_T=\frac{4 a c T^3}{3\kappa C_P \rho^2}.
\label{eq:kappamu}
\end{equation}
In this last equation $a$ stands for the radiation density constant, $c$ is the speed of light and $\kappa$ the Rosseland mean opacity.

\citet{traxler-garaud-stellmach11} solved Eqs. (\ref{eq:trax6}-\ref{eq:trax9}) in a triply periodic box of size $(L_x, L_y, L_z)$ and carried out simulations for moderately low values of the Prandtl number and diffusivity ratio of order 
$O(10^{-2})$. 

As a result of numerical experiments,  it turns out that the turbulent compositional transport by fingering convection follows a simple law for the diffusion coefficient, namely
\begin{equation}
D_\mu = 101 \sqrt{\kappa_\mu \nu}\, e^{-3.6r}(1-r)^{1.1},
\label{eq:dif-coef-trax}
\end{equation}
where $r=(R_0-1)/(\tau^{-1}-1)$ and $\nu$ is the total viscosity given by the sum of the molecular and radiative viscosities \citep{denissenkov10} 
\begin{equation}
\nu = \nu_\mathrm{mol} + \nu_\mathrm{rad}
\label{eq:total-viscosity}
\end{equation}
with
\begin{equation}
\nu_\mathrm{rad}=\frac{4 a T^4}{15 c \kappa \rho^2}
\label{eq:rad-viscosity}
\end{equation}
and
\begin{equation}
\nu_\mathrm{mol}\equiv \kappa_\mu =1.84 \times 10^{-17} (1+7 X)\frac{T^{5/2}}{\rho}, \qquad [\mathrm{cm}^2 \mathrm{s}^{-1}]
\label{eq:mol-viscosity}
\end{equation}
where $X$ is the hydrogen mass fraction.
Based on the asymptotic behavior shown by their results, \citet{traxler-garaud-stellmach11} suggest the possibility of applying Eq. (\ref{eq:dif-coef-trax}) to the more extreme  astrophysical regime, provided Pr is of the order of $\tau$.

%
\section{Numerical simulations}

In order to study the effects of thermohaline instability in low-mass giant stars we performed simulations using a one-dimensional evolution code \citep[LPCODE,][]{althaus-serenelli-panei-et-al05} incorporating GNA's convection theory to compute the mixing rates of the different stability regimes defined by this formalism. 
We adopted in our numerical experiments the following choice for the parameters: $A=0$, $F=0$, $D=3K/(\rho C_P l^2)$, $\alpha=1$, $\phi=1$ and $l=1.35$ (approximately equivalent to a mixing length parameter of 1.61 in the usual \citealt{kippenhahn-weigert90} prescription), and implemented the same nuclear reaction network as described at length by \citet{althaus-serenelli-panei-et-al05}.

\begin{figure}
\resizebox{\hsize}{!}{\includegraphics{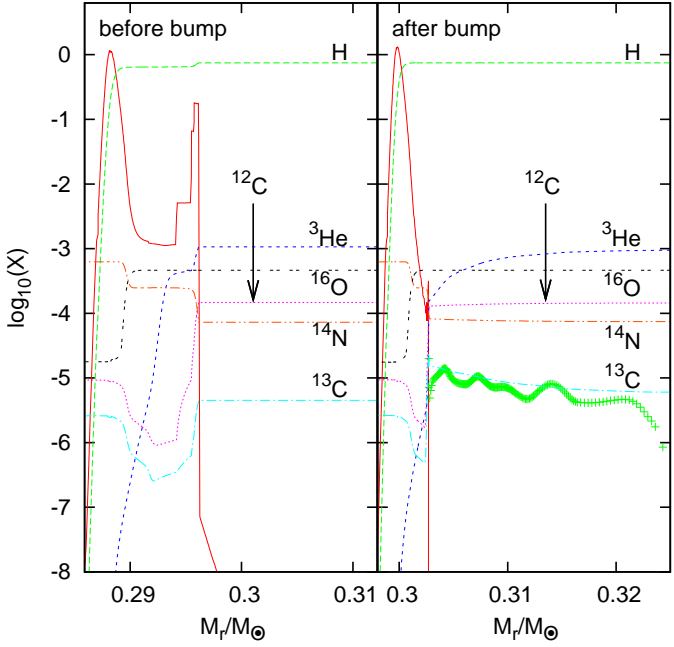}}
\caption{Same as Fig. \ref{figure2} for an artificially increased diffusion coefficient (see text).} 
\label{fig-Dx1000}
\end{figure}

We computed stellar models of 0.9 $M_{\sun}$, 1.3 $M_{\sun}$ and 1.6 $M_{\sun}$, each one with three different initial metallicities, namely Z=$3.17\times 10^{-4}$, 
$1\times 10^{-3}$ and $6.32\times 10^{-3}$, and let them evolve from the main sequence until after the luminosity bump. For each model, we paid special attention to the detailed stellar structure of the region between the HBS and the base of the convective envelope, contrasting the situation given before the star enters the luminosity bump region and after that stage of its evolution. The results obtained in all cases were qualitatively very similar, thus we show here just one case, namely the 0.9 $M_{\sun}$, 
Z=$1\times 10^{-3}$ model, which is representative of what happens to the others.
Figure \ref{figure2} shows the abundance profiles of some elements and the mean molecular weight gradient for the 0.9 $M_{\sun}$, Z=$1\times 10^{-3}$ model, before and after the luminosity bump. Before the luminosity bump (left panel in Fig. \ref{figure2}), the mean molecular weight gradient $\nabla_\mathrm{\mu}$ shows two peaks, corresponding to the hydrogen burning shell (left peak) approaching the molecular weight discontinuity (right peak) left behind by the first dredge-up. When the hydrogen burning shell reaches the discontinuity, the reaction $^3$He($^3$He,2p)$^4$He produces a molecular weight inversion in the external tail of the HBS, destabilizing the region and producing thermohaline convection. The destabilized zone (shown by plus signs in the right panel of Fig. \ref{figure2}) never reaches the convective envelope in our simulations, being both regions separated by a radiative zone that prevents any change in the surface composition of the star. This result clearly differs from the main result presented by CZ07 when the slender finger geometry of \citet{ulrich72} was adopted. This should not come as a surprise as our assumption of a unique mixing length in GNA theory is far from a slender finger geometry. In fact, our results are consistent with those of CZ07 
when blobs, rather than slender fingers, are assumed. As shown by CZ07, different blob/finger geometries can affect the diffusion coefficient by more than two orders of magnitude. In this connection, we performed additional simulations artificially increasing GNA's diffusion coefficient of thermohaline unstable layers by a factor of $10^3$ in order to test the eventual relation between the more rapid mixing rate and the surface abundance variations. Figure \ref{fig-Dx1000} shows the abundance profile of the same elements included in Fig. \ref{figure2} as well as the run of the molecular weight gradient in the region comprised by the HBS and the base of the convective envelope, for this new experiment. Now the thermohaline zone expands outwards (in mass) occupying all the former radiative region that separated it from the convective envelope. The contact between both convective regions allows for the non-canonical extra mixing to take place, thus modifying the photospheric chemical composition after the luminosity bump. 

Given the contrast between the diffusion coefficients computed as given by the GNA and those reported by CZ07, we decided to perform further simulations combining the GNA theory with other prescriptions used to estimate the diffusion coefficient. 
Only very few different prescriptions for the computation of the diffusion coefficient in thermohaline unstable regions exist in the literature. 
\citet{ulrich72} was the first to derive an expression for the turbulent diffusivity produced by that instability, whereas \citet{kippenhahn-ruschenplatt-thomas80} extended previous works to the case of a non-perfect gas. The linear theory used by these early works yield a solution for the diffusion coefficient that is proportional to the square of the unknown aspect ratio (length to diameter) of the fluid elements, an issue that is still a matter of debate. Indeed, the diffusion coefficients may differ by about two orders of magnitude depending on the form factor adopted by different authors. Thus, the implementation of the linear theory has the drawback of containing a high intrinsic uncertainty. On the other hand, more recently, \citet{traxler-garaud-stellmach11} successfully derived empirically determinated transport laws for thermohaline unstable regions by means of three-dimensional simulations performed at parameter values approaching those relevant for astrophysics. This represents an alternative and more physically sound approach which helps us to
avoid the problems of the classical linear theory and supplies an independent way to address the question of the actual role of mixing in thermohaline unstable regions.

Since GNA convection theory may be implemented to determine the regime of energy transport of any layer with the advantage of leaving the computation of the diffusion coefficient as an independent task which might subscribe to different prescriptions, we decided to study the system's response combining GNA formalism with two independent recipes. On the one hand, we computed the thermohaline diffusion coefficient by means of the expression obtained by \citet{kippenhahn-ruschenplatt-thomas80} 
\begin{equation}
D_\mathrm{K}= \alpha_\mathrm{th} \frac{3K}{2\rho C_P} 
\frac{\frac{\phi}{\delta} \nabla_\mathrm{\mu} }{(\nabla-\nabla_\mathrm{ad})},
\label{eq:difcoef-kippenhahn}
\end{equation}
where $\alpha_\mathrm{th}$ is a efficiency parameter which depends on the geometry of the fluid elements, $\rho$ is the density, $K=4acT^3/(3\kappa\rho)$ the thermal conductivity, and $C_P=(\mathrm{d}q/\mathrm{d}T)_P$ the specific heat capacity. We set $\alpha_\mathrm{th}=2$, which roughly corresponds to the prescription of \citet{kippenhahn-ruschenplatt-thomas80}. On the other hand, we computed diffusion coefficients adopting the \citet{traxler-garaud-stellmach11} empirical law given by Eq. (\ref{eq:dif-coef-trax}).

\begin{figure}
\resizebox{\hsize}{!}{\includegraphics{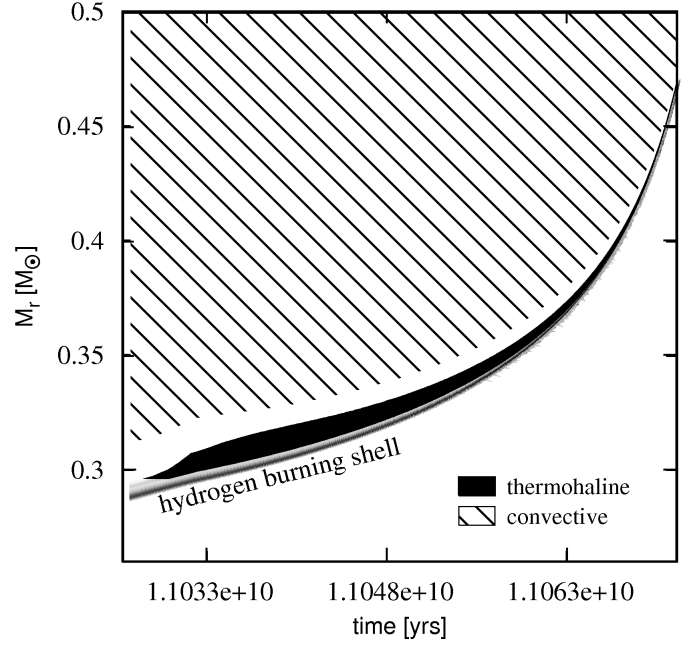}}
\caption{Evolution of the region between the HBS and the convective envelope, when the prescription of \citet{kippenhahn-ruschenplatt-thomas80} is adopted to compute the diffusion coefficient in the thermohaline zone. Time interval spans from the instant when the stars luminosity reaches $L\approx 96$ (i.e., before the luminosity bump) until $L\approx 1826 $, close to the top of the RGB. The figure corresponds to the 0.9 $M_{\sun}$, Z=$1\times 10^{-3}$, model.}
\label{fig:evol-layers-kip}
\end{figure}

\begin{figure}
\resizebox{\hsize}{!}{\includegraphics{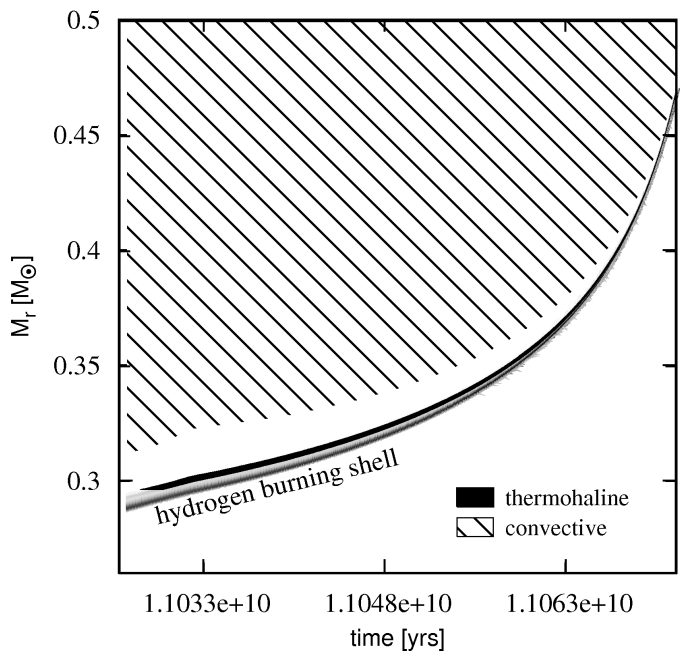}}
\caption{Same as Fig. \ref{fig:evol-layers-kip} but when the prescription of \citet{traxler-garaud-stellmach11} is adopted to compute the diffusion coefficient in the thermohaline zone. The figure corresponds to the 0.9 $M_{\sun}$, Z=$1\times 10^{-3}$, model.}
\label{fig:evol-layers-trx}
\end{figure}

Thus, we performed a few additional simulations for the 0.9 $M_{\sun}$, Z=$1\times 10^{-3}$, and 1 $M_{\sun}$, Z=$0.02$, sequences in order to investigate the response of the system when we solely vary the recipe to compute diffusion coefficients. Fig. \ref{fig:evol-layers-kip} shows the evolution of the thermohaline region along the RGB when the prescription of \citet{kippenhahn-ruschenplatt-thomas80} is adopted. Note that the convective envelope never enters into contact with the thermohaline region. Consequently, for this model and mixing treatment, the photospheric abundances of the star remain constant throughout this phase. A similar behavior is shown by Fig. \ref{fig:evol-layers-trx}, corresponding to the implementation of the recipe of  \citet{traxler-garaud-stellmach11}. In this case, the thermohaline zone is much more narrow than before. We will see in the next section that this fact is in close relation with the magnitude of the diffusion coefficients computed using different prescriptions. Finally, it is worth noting that in our 1 $M_{\sun}$ , Z=$0.02$, sequence, we did not find any contact between the bottom of the convective envelope and the thermohaline region. This result is at variance with the simulations presented by \citet{cantiello-langer10} which showed that this contact occurred in 1 $M_{\sun}$ mass stars even for the prescription of \citet{kippenhahn-ruschenplatt-thomas80} with $\alpha_\mathrm{th}=2$. We suspect that this different behavior may be due to the different microphysics assumed in both stellar codes.

%
\section{Summary and discussion}
We have studied the impact of thermohaline mixing in red giants close to the luminosity bump in the light of two non-standard and physically sounding mixing prescriptions: the GNA and the \citet{traxler-garaud-stellmach11} prescription. To the best of our knowledge, this is the first time that the empirically based thermohaline mixing prescription of \citet{traxler-garaud-stellmach11} is tested in the context of detailed evolutionary simulations.  In the case of the double diffusive mixing length theory of \citet{grossman-narayan-arnett93} it has allowed us to include thermohaline mixing in a consistent way with the other unstable regimes that are possible when $\nabla_\mathrm{\mu}\ne 0$, selfconsistently solving the temperature gradients and turbulent mixing rates.

For the sake of completeness let us mention that in the case of GNA theory, we find the thermohaline mixing efficiency to be very similar to that of \citet{kippenhahn-ruschenplatt-thomas80}. In fact our computations show that at almost all layers the value predicted by \citet{grossman-taam96} is $D_\mathrm{GNA}\sim D_{\rm Kip}/6$, a difference that is just a consequence of different choices in the adimensional coefficients of both prescriptions. The similarities between these two prescriptions should not come as a surprise as the GNA theory is a sophisticated version of the mixing length theory but still relies on a very similar picture than the standard MLT, in which \citet{kippenhahn-ruschenplatt-thomas80} prescription is based on. We consider our results as an actual validation of the GNA theory for the cases where \citet{kippenhahn-ruschenplatt-thomas80} prescription is applicable.

Both \citet{traxler-garaud-stellmach11} and \citet{grossman-narayan-arnett93} prescriptions have identified thermohaline mixing to develop in RGB stars close to the luminosity bump, in agreement with all previous work that have adopted more simplified approaches \citep{charbonnel-zahn07,cantiello-langer10}.

\begin{figure}
\resizebox{\hsize}{!}{\includegraphics{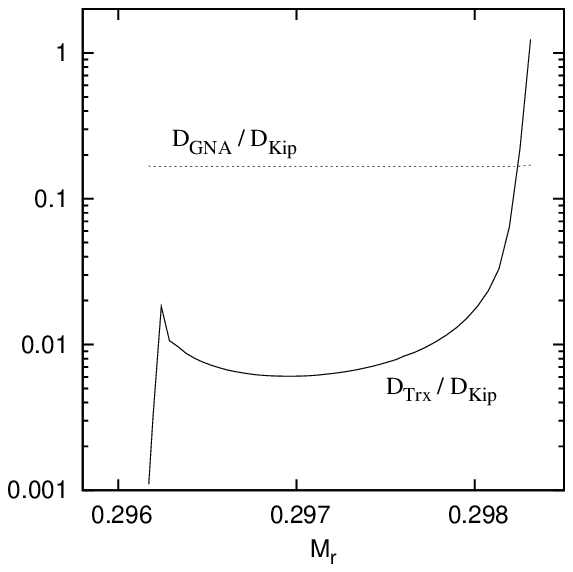}}
\caption{Relation between diffusion coefficients computed using the prescriptions of \citet{kippenhahn-ruschenplatt-thomas80}, $D_\mathrm{Kip}$, \citet{traxler-garaud-stellmach11}, $D_\mathrm{Trx}$, and \citet{grossman-narayan-arnett93}, $D_\mathrm{GNA}$.
}
\label{fig:dif-compare}
\end{figure}

However, in agreement with \citet{denissenkov10} and \citet{traxler-garaud-stellmach11} suggestions, our full evolutionary calculations confirm that thermohaline mixing is not efficient enough for fingering convection to reach the bottom of the convective envelope of red giants. Thus, no changes in the surface chemical abundances of red giants are obtained when either \citet{traxler-garaud-stellmach11} or \citet{grossman-narayan-arnett93} prescriptions are adopted. Interestingly enough, as the value of $(\nabla-\nabla_\mathrm{ad})/\nabla_\mathrm{\mu}$ in the thermohaline zone is $(\nabla-\nabla_\mathrm{ad})/\nabla_\mathrm{\mu}\sim 10^{3}...\sim 10^{4}$ it falls in a regime in which the standard prescription of \citet{kippenhahn-ruschenplatt-thomas80} strongly overestimates the thermohaline mixing efficiency \citep[see Fig. 3 of][]{traxler-garaud-stellmach11}. As can be seen in Fig. (\ref{fig:dif-compare}) the standard prescription is $\sim 100$ to 1000 times more efficient than the empirical \citet{traxler-garaud-stellmach11} law. However, we know from \citet{cantiello-langer10} that the standard prescription is still not enough to account for the surface abundances of RGBs. Thus, in order to allow contact between the thermohaline region and the convective envelope, the diffusion coefficient should be about 4 orders of magnitude higher than predicted by realistic thermohaline transport laws \citep{denissenkov10,traxler-garaud-stellmach11}. {\bf Since hydrodynamics codes have shown to be consistent, yielding similar results between different implementations, they should be trusted in the physical regime studied (Pr $\,\geq 10^{-2}$) which, due to computational limitations, is not the actual astrophysical regime (Pr$ \sim 10^{-6}$). While the prescriptions used here still relies on an asymptotic scaling, it seems difficult that the diffusion coefficients could be off by this much.} Thus, we can conclude that thermohaline mixing alone is {\bf very unlikely to be} the explanation for the chemical abundance anomalies of red giants.

\begin{acknowledgements}
We thank the valuable comments and suggestions of our referee which helped us to strongly improve the original version of this paper.
We acknowledge financial support from PIP-112-200801-00904 by CONICET and PICT-2006-00504 by ANCyT.
\end{acknowledgements}

\bibliographystyle{aa} 
\bibliography{biblio-base} 
%
\appendix
\label{sec:apendix}
\section{Solving GNA's equations}
GNA theory of convection provides us with a set of equations that have to be solved in order to find the values of the temperature gradient $\nabla$ and the turbulent velocity $\sigma$ of the system in regions of different energy transport regime. In practice, this means to solve equations (\ref{eqnsigma}) and (\ref{eqnflux}) simultaneously. Being it impossible to express any of these variables in terms of the other we followed an iterative procedure, adopting the Newton-Raphson method to this end. In order to avoid eventual numerical instabilities associated to the divergence of Eq. (\ref{eq:wtheta}) when the denominator becomes small, we elementary transformed equation (\ref{eqnflux}) by multiplying it by that denominator, and rearranging the flux conservation equation we obtain  
\begin{equation}
  X^2 + a_1(\sigma,Y,X_\mathrm{Rad})\, X + a_2 (\sigma,Y,X_{Rad}) = 0,
\label{eq:fluxnew}
\end{equation}
where we adopted the following nomenclature
\begin{eqnarray}
& & X = g\alpha (\nabla-\nabla_\mathrm{ad})/H_\mathrm{p}, \\
\label{eq:X}
& & X_{Rad} = g\alpha (\nabla_\mathrm{Rad}-\nabla_\mathrm{ad})/H_\mathrm{p}, \\
\label{eq:XRad}
& & Y  =  g\phi \nabla_\mathrm{\mu}/H_\mathrm{p}.
\label{eq:Y}
\end{eqnarray}
$a_1$ and $a_2$ are the coefficients of the quadratic equation in $X$ (i.e., $\nabla$), which depend explicitly on the turbulent velocity $\sigma$, the composition gradient $\nabla_\mathrm{\mu}$ and the total radiation gradient $\nabla_\mathrm{Rad}$.

Thus, given a set of diffusion rates of heat ($D$), composition ($F$) and momentum ($A$), and once $X_\mathrm{Rad}$ and $Y$ are known, we first determine if the total radiation might be transported in a not convective way. If radiative transport is insufficient, convection has to carry some fraction of the energy flux, and thus we  start the iterative procedure above mentioned. We adopt an initial (guess) value for $X$ and solve Eq. (\ref{eqnsigma}) for $\sigma$. As stated before, the system will seek out the most turbulent equilibrium state, thus we solve both cubic equations and pick up the largest positive root. The adopted values for $X$ and $\sigma$ are then introduced in Eq. (\ref{eq:fluxnew}) and Newton-Raphson method is used in order to find the correction to be applied to $X$. Iterating this process it is possible to obtain the values of $X$ and $\sigma$ that satisfy equations (\ref{eqnsigma}) and (\ref{eq:fluxnew}). Numerical experiments have shown that $X=X_\mathrm{Rad}$ is a good starting value for the Newton-Raphson process, while other choices resulted in false roots found by the algorithm.

Finally, it is worth mentioning that factors in brackets in Eq. (\ref{eqnsigma}) are cubic in $\sigma$ and, since both cubics are different, the conditions that separate the real roots region from the one real plus two complex conjugate roots region are also different. Despite this difference, for the stellar astrophysics case we have that $D\gg A$ and $D\gg F$, and both conditions tend to the same curve, thus being unnecessary in practice to compute both limiting curves.

\end{document}